\documentclass[%
 reprint,
 amsmath,amssymb,
 aps,
floatfix,
]{revtex4-2}

\usepackage{verbatim} 

\usepackage{graphicx}
\usepackage{float}
\usepackage{dcolumn}
\usepackage{bm}

\usepackage{color}
\usepackage{xcolor}

\begin{document}


\title{Schramm-Loewner evolution in 2d rigidity percolation}

\author{Nina Javerzat}

 \email{njaverza@sissa.it}
\affiliation{%
 SISSA and INFN Sezione di Trieste, via Bonomea 265, 34136, Trieste, Italy
}%

\date{\today}

\begin{abstract}
Amorphous solids may resist external deformation such as shear or compression while they do not present any long-range translational order or symmetry at the microscopic scale. Yet, it was recently discovered that, when they become rigid, such materials acquire a high degree of symmetry hidden in the disorder fluctuations: their microstructure becomes statistically conformally invariant. In this Letter we exploit this finding to characterise the universality class of central-force rigidity percolation (RP), using Schramm-Loewner Evolution (SLE) theory. We provide numerical  evidences that the interfaces of the  mechanically stable structures (rigid clusters), at the rigidification transition, are consistently described by SLE$_\kappa$, showing that this powerful framework can be applied to a mechanical percolation transition.
Using well-known relations between different SLE observables and the universal diffusion constant $\kappa$, we obtain the estimation $\kappa\sim2.9$ for central-force RP. This value is consistent, through relations coming from conformal field theory, with previously measured values for the clusters' fractal dimension $D_f$ and correlation length exponent $\nu$, providing new, non-trivial relations between critical exponents for RP. These findings open the way to a fine understanding of the microstructure in other important classes of rigidity and jamming transitions.
\end{abstract}

\maketitle

\textit{Introduction} -- Predicting the mechanical behaviour of amorphous media, which have no long-range structural order, remains a challenge. Significant progress has been made in the last years to unify the description of disordered solids \cite{silke2009,degiuli2018,bulbul2020} like grain packings \cite{bulbul2022} or gels \cite{emanuela2023}. A lot remains to be understood though: indeed very diverse materials, from molecular glasses \cite{Thorpe2000} and gels \cite{mehdi2019,mehdi2022} to fibers \cite{Head2003,Broedersz2011,Gurmessa2019} and  living tissues \cite{Petridou2021,Jackson2022,Lenne2022} undergo a sudden transition between a fluid and a rigid state as an external parameter --eg. the volume fraction of constituents-- is varied. Such ability proves crucial for instance in biological processes like embryogenesis \cite{Petridou2021}. However, a quantitative understanding of the behaviour of materials near rigidity transitions remains generally elusive.  Can we find a unified description of rigidity transitions, predicting both the structural and mechanical properties of amorphous media at and close to their solidification point ? \\

Rigidity percolation (RP) \cite{Thorpe1983} plays a key role in this context, as a simple framework to model the emergence of mechanical stability in a disordered network: as the density of microscopic components is increased, they form structures, rigid under deformation which eventually percolate the whole system, ensuring resistance to macroscopic constraints. The sudden change of mechanical state --from fluid to rigid, gets recasted in the well-known language of percolation theory and controlled by the percolation order parameter, the probability that a network component belongs to the percolating rigid cluster.
RP thus represents a general framework to model the solidification of amorphous media, and has been successfully applied to understand the mechanical behaviour of the materials listed above \cite{Thorpe2000,mehdi2019,mehdi2022,Head2003,Broedersz2011,Gurmessa2019,Petridou2021,Jackson2022,Lenne2022}. 
A considerable amount of work has been devoted to the onset of rigidity 
\cite{deGennes1976,Feng1984,Kantor1984,JacobsThorpe1995,Moukarzel1997,Duxbury1998,Moukarzel1999,Head2003,Plischke2007,Broedersz2011,xiaoming2015,ellenbroek2015,silke2016,Liarte_2016,Liu2019,mehdi2019,nina2022} corresponding in many cases to second-order transitions defining new universality classes. Their characterisation remains however limited to the numerical measurement of the main critical exponents, eg. $\nu$ and $\beta$ controlling respectively the behaviour of the correlation length and the order parameter at the transition.

RP is notably found to be distinct from connectivity percolation (CP) \cite{JacobsThorpe1995,Liu2019}, which is simply defined by the emergence of a (non-rigid) percolating cluster 
and features a rich and well-studied critical behaviour, with clusters becoming random fractals \cite{AharonyBook}. RP adds the mechanical dimension to the problem making it intrinsically non-local, where the removal of a bond may destroy rigidity over a large region \cite{JacobsThorpe1995,Moukarzel1999}. RP transitions remain therefore much less understood: the computation of the critical exponents, the possible connections with CP and the possible unification of different RP universality classes are long-standing questions \cite{deGennes1976,Feng1984,Kantor1984,Moukarzel1999,Plischke2007,Liu2019,nina2022} still not fully resolved to date.

\begin{figure}
    \centering
    \resizebox{\columnwidth}{!}{
    \includegraphics[width=\linewidth]{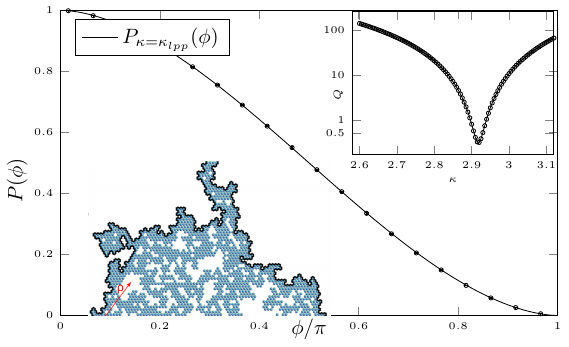}}
    \caption{Main: Left-passage probability of spanning perimeters and the prediction (\ref{eq:LPP}) for $\kappa=\kappa_{\mathrm{LPP}}$. Top-right inset: Weighted mean square deviation $Q$ given by (\ref{eq:Q}). Bottom-left inset: Example of a rigid cluster (blue) and its left and right hulls (black thick lines). The right hull passes to the left of the point located at $z=\rho e^{i\phi}$ from its origin, marked as a red vector.}
    \label{fig:lpp}
\end{figure}

Recently, deep connections between CP and RP were uncovered. Liu et al \cite{Liu2019} studied minimal rigidity percolation (MRP), where the number of network's degrees of freedom matches exactly the number of constraints, and argued that it falls in the CP universality class.
Even more recently \cite{nina2022} it was found, based on a numerical study of a particular set of observables, that central-force RP clusters are conformally invariant at the critical point. This opens the possibility to describe RP in the framework of conformal field theory (CFT), in which scaling limits of observables are expressed as appropriate correlation functions \cite{Cardy_notes}. Conformal invariance --the invariance of the statistical properties under local rescalings of the system, appears in many critical phenomena \cite{Polyakov1970,DiFrancesco} but its emergence in RP is quite remarkable given the high non-locality of the problem. Conformal symmetry has been used with great success to predict the universal properties of critical points \cite{Belavin1984,BPZ84,Friedan1984,Poland2016}, and notably allowed to determine exactly the critical exponents and other universal quantities in percolation \cite{DuplantierSaleur1987,diFrancesco1987,Cardy_1992,Delfino_2011,Picco2016,He2020}. For rigidity transitions too, exploiting conformal invariance can give access to the fine universal properties of the medium's microstructure. Besides achieving a more complete characterisation of the RP universality classes and their potential interconnections, this is also especially important regarding the prediction of the mechanical aspects of rigidity transitions. Indeed, the interplay between microstructure and elastic behaviour at and near criticality has been recently highlighted in particulate materials \cite{mehdi2022}; although not systematically understood \cite{Head2003}, it reveals that characterising the microstructure near the transition --where structural heterogeneities develop at all length scales, is crucial to understand the mechanical response.\\\\
\indent In this Letter we investigate the RP universality class using Schramm-Loewner Evolution (SLE) theory, a powerful framework to exploit conformal symmetry, which proved notably extremely successful to understand connectivity percolation \cite{Cardy2005}. We focus for simplicity on the case of purely central forces, and consider the interfaces separating rigid clusters from floppy regions at criticality,  known as the complete perimeters or hulls \cite{AharonyBook} of rigid percolating clusters (see inset of figure \ref{fig:lpp}). In CP \cite{Schramm2000,Camia2007}, as well as in diverse critical phenomena \cite{Amoruso2006,Bernard2007,Smirnov2007,Gamsa_2007,Santachiara2008,Jacobsen2009,Rohde2011,Caselle_2011}, the scaling limit of random interfaces are very interesting objects, as their probability measure possesses two remarkable properties, conformal invariance and a domain Markov property (DMP),
which together yield powerful results about the universal class through the theory of SLE (see eg. \cite{Kager2004,Cardy2005,Bauer_review} for introductions to the topic). The basic idea of SLE \cite{Schramm2000} is that a curve $\gamma_t$ in a planar domain, parametrised by the time $t$ can be equivalently encoded in a real function --the driving-- $\xi_t$ through a series of conformal maps $g_t$, such that the tip of the growing curve is $\gamma_t=g_t^{-1}(\xi_t)$.
For a conformally invariant curve $\xi_t = \sqrt{\kappa}B_t$ with $B_t$ a standard Brownian motion and $\kappa$ a universal parameter known as diffusion constant \cite{Schramm2000}. As a consequence of such remarkable equivalence, the scaling limit of the curve is determined by the value of $\kappa$ only and many observables can be predicted exactly. This provides a classification of physically relevant random curves according to their value of $\kappa$ \cite{Wieland2003,Bernard2006,Amoruso2006,Bernard2007,Boffetta2008,Saberi2009,Saberi2010,Herrmann2012,Pose2014,Giordanelli2016,deCastro2018}.
The central charge $c$ of the corresponding CFT is related to $\kappa$ \cite{Bauer2002}:
\begin{equation}\label{eq:c}
    c = \frac{(8-3\kappa)(\kappa-6)}{2\kappa}
\end{equation}
$c$ encodes how conformal symmetry manifests itself  in the system 
and gives eg. the change in free energy when conformally transforming the system's geometry \cite{blote86,affleck,Cardybook}. Its value is needed to compute physical observables and represents therefore a crucial information about the universality class.\\

As we will argue, interfaces in RP are numerically consistent with SLE. Besides giving access to more universal observables, this finding reveals the existence of a non-trivial algebraic structure in the underlying CFT, through the existence of a so-called degenerate field \cite{Bauer2002,Cardy2005}. This places the CFT of rigidity percolation in a more restricted class of field theories, to which connectivity percolation also belongs, and allows to express RP critical exponents solely in terms of $\kappa$. \\

\textit{Model and methods} -- 
We study central force rigidity percolation on the site-diluted model with local correlations introduced in \cite{mehdi2019}: we populate randomly the sites of a triangular lattice of $L_1$ columns and $L_2$ rows with probability $p(\mathrm{site}) = (1-\tilde{c})^{6-N_n(\mathrm{site})}$, where $N_n$ is the number of occupied nearest neighbours of that site. The short-range correlations introduced by taking $\tilde{c}>0$ decrease the rigidity percolation threshold \cite{mehdi2019}, allowing to generate critical configurations of lower density hence reducing the computation time, without affecting the long-range behaviour so that universal quantities keep the same values as in the uncorrelated case \cite{mehdi2019,nina2022}. We take $\tilde{c}=0.3$, corresponding to a rigidity threshold $p_c\sim 0.657$ \cite{mehdi2019}. Rigid clusters are identified by the so-called Pebble Game \cite{JacobsThorpe1995} which, by efficient constraint counting allows to test the mutual rigidity of the bonds connecting occupied sites. We select the configurations where at least one rigid cluster percolates from bottom to top and construct, for each percolating cluster, its complete perimeter defined on the dual (hexagonal) lattice
as shown in the inset of figure \ref{fig:lpp}; the hull of the rigid cluster (in blue) is highlighted as a dark line.

We stress that such unidirectionally percolating clusters have the same fractal structure as the cross-percolating ones, which ensure mechanical stability under global deformation. It is therefore sufficient, to study the random geometry of RP, to restrict to the former type of clusters which are most adapted to the SLE analysis as detailed below.

\noindent We use the strip geometry, namely a large enough aspect ratio $L_1/L_2=4$, with periodic horizontal boundary conditions and open vertical boundary conditions, and systems as large as $L_2=1024$; sample averages, denoted $\mathbb{E}\left[\cdots\right]$, are performed over $N\geq 40\,000$ curves. Curves start from a point on the lower boundary and grow until they hit the upper boundary, and we therefore choose the framework of dipolar SLE \cite{Bauer_2005}, consistent with such a setup. In that case the conformal maps $g_t$ satisfy the Loewner equation \cite{Bauer_2005}
\begin{equation}\label{eq:Loewner}
    \frac{dg_t(z)}{dt} = \frac{\pi/L_y}{\tanh\left[\frac{\pi}{2L_y}\left(g_t(z)-\xi_t\right)\right]},\quad g_0(z)=z
\end{equation}
where $L_y=L_2 \sqrt{3}/2$ is the width of the strip.

In the following sections we: i) establish that the complete perimeters of the spanning rigid clusters are SLE$_\kappa$ by showing that the statistics of their driving functions is compatible with a Brownian motion, and ii) obtain independent numerical estimates of the value of the diffusivity $\kappa$ from the measurement of universal properties of the curves: their fractal dimension, winding angle and left-passage probability. The results are summarised in table \ref{tab:kappas}.\\


\textit{Driving function} -- To extract numerically the driving $\xi_t$ of a given curve, the idea is to solve equation (\ref{eq:Loewner}) for each short time interval $\delta t$ on which $\xi_t$ is approximated as constant, obtaining the slit-map $g_t$ (see eg. \cite{Kennedy2008,Cardy2005}). 
Then, for each lattice curve $\left\{z_0^0=0,z_1^0,\cdots,z_l^0\right\}$ of length $l$ starting at the origin, we compute iteratively the Loewner times $t_j$ and the driving function $\xi_{t_j}$ by successive applications of $g_{t_j}$, such that $t_0=0$, $\xi_{t_0}=0$, and at each step the sequence $\left\{z_j^{j-1},\cdots,z_l^{j-1}\right\}, j\geq1$ is mapped to the reduced sequence $\left\{z_{j+1}^j = g_{t_j}(z_{j+1}^{j-1}),\cdots,z_l^j=g_{t_j}(z_l^{j-1})\right\}$. For dipolar SLE the $g_{t_j}$ are given by \cite{Bernard2007}
\begin{equation}
    g_{t_j}(z) = \xi_{t_j}+\frac{2L_y}{\pi}\cosh^{-1}\left[\frac{\cosh\left[\frac{\pi}{2L_y}(z-\xi_{t_j})\right]}{\cos \Delta_j}\right]
\end{equation}
where $\xi_{t_j} = \mathrm{Re}(z_j^{j-1})$, $t_j = t_{j-1} -2\left(\frac{L_y}{\pi}\right)^2\log\left(\cos \Delta_j\right)$, $\Delta_j = \frac{\pi}{2L_y}\mathrm{Im}(z_j^{j-1})$ and we compute the complex inverse hyperbolic cosine as
$\cosh^{-1} z = \log\left[z+i\sqrt{-z^2+1}\right]$. Each curve of length $l$ is unzipped in this way for $l/2$ steps, and yields an instance of the driving function $\xi_t$ at sample-dependent, non-equally spaced Loewner times $t_0,\cdots, t_{l/2}$. We linearly interpolate each instance of the driving function to have all instances defined for a same, equally spaced, time sequence.

\begin{figure}
    \centering
    \resizebox{\columnwidth}{!}{
    \includegraphics[width=\linewidth]{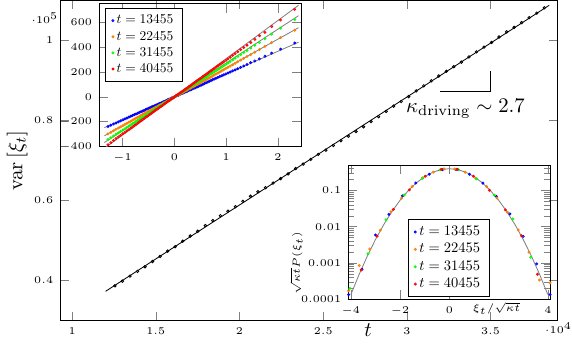}}
    \caption{Main: Variance of the driving function $\xi_t$ as a function of time. Bottom-right inset: Rescaled PDF of $\xi_t$ at 4 different times; in grey the Gaussian distribution. Top-left inset: Quantile-quantile plots of $\xi_t$ vs $\mathcal{N}(0,1)$ at 4 different times. Grey lines have slopes $\sqrt{\kappa_{\mathrm{driving}}\,t}$.}
    \label{fig:xi}
\end{figure}

Loewner's equation (\ref{eq:Loewner}) holds for generic curves, and to claim that the rigid perimeters are indeed SLEs we must ensure that the extracted driving function is a Brownian motion. The insets of figure \ref{fig:xi} show the distributions of $\xi_t$ at different times and the corresponding quantile-quantile plots, both consistent with Gaussian distributions.
However, as studied by Kennedy \cite{Kennedy2008}, gaussianity at fixed times alone is not an accurate test as it is passed by non-SLE processes as well, and one must test also the independence of the driving function's increments. Following \cite{Kennedy2008} we pick $n$ equally spaced times $0<t_1<\cdots<t_{n}$ and define the $n$ increments $X_j \equiv \xi_{t_j+\delta} - \xi_{t_{j}}$. We set $\delta=5$ but choosing a different value does not affect our conclusions as long as $\delta\ll \tau\equiv t_{j+1}-t_j$ (see also \cite{Bernard2007,Pose2014,deCastro2018} where the correlation between two such consecutive increments is seen to decay for $\tau\gg\delta=1$). The joint distribution of $\left(X_1,\cdots,X_n\right)$ is tested by defining $m=2^n$ cells, each corresponding to the possible sign sequence of $\left(X_1,\cdots,X_n\right)$, counting the number $O_j$ of samples falling in each cell and comparing with the expected value, $E_j=N/2^n$ for independent and Gaussian distributed variables. To this end one defines
\begin{equation}\label{eq:chi}
    \chi^2 = \sum_{j=1}^m \frac{\left(O_j-E_j\right)^2}{E_j}
\end{equation}
and computes the associated p-value. 
Taking increasing numbers $n$ of increments, with $t_k\in[9.10^3,29.10^3]$, we find $p_{n=5}=0.95$, $p_{n=7}=0.78$, $p_{n=9}=0.81$.

These p-values are not small, indicating that one cannot reject the hypothesis that $\xi_t$ is a Brownian motion. This leads to our first main result, that the statistics of the hulls of rigid RP clusters are consistent with SLE. 


The driving function gives straightforwardly a first estimate of the diffusion constan, since by definition $\mathrm{var}\left[\xi_t\right] = \kappa\,t$. The main plot of figure \ref{fig:xi} shows the variance of $\xi$ as function of time, and a fit gives the value $\kappa_{\mathrm{driving}}$ reported in table \ref{tab:kappas}.

\begin{table}[H]
    \centering
    \begin{tabular}{c|c|c|c}
         $\kappa_{\mathrm{fractal}}$    & $\kappa_{\mathrm{winding}}$    & $\kappa_{\mathrm{LPP}}$  & $\kappa_{\mathrm{driving}}$\\ \hline
         $2.84(8)$        &$2.88(5)$       &  $2.91(1)$             & $2.7(1)$
    \end{tabular}
    \caption{Values of $\kappa$ obtained from the fractal dimension, winding angle, left-passage probability and zipper algorithm, as described in the text.}
    \label{tab:kappas}
\end{table}

It was observed however that the estimation of $\kappa$ from the driving function comes with non-negligible error \cite{Pose2014}.
In the next sections we therefore use standard results on SLE to obtain independent and more accurate estimations of $\kappa$ from more directly measurable quantities.\\

\begin{figure}
    \centering
    \resizebox{\columnwidth}{!}{
    \includegraphics[scale=1,width=\linewidth]{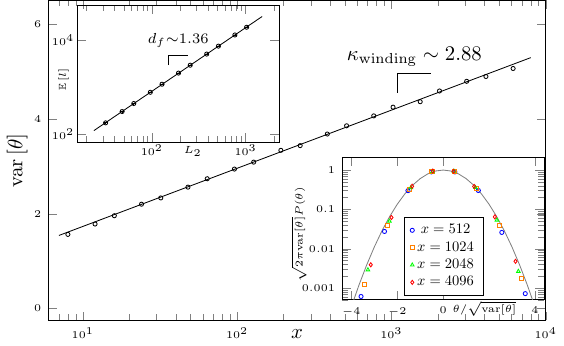}}
    \caption{Main: Variance of the winding angle $\theta$ as function of the distance $x$ along the curve. Top-left inset: Mean length $\mathbb{E}\left[l\right]$ of the curves as function of the system height $L_2$. Bottom-right inset: Probability distribution of the winding angle at different distances $x$; in grey the Gaussian distribution.}
    \label{fig:w}
\end{figure}

\textit{Fractal dimension} -- For a spanning curve of lattice length $l$ the fractal dimension can be defined as $l\sim L_2^{d_f}$. For SLEs, $d_f$ is related to the diffusivity by $d_f=1+\kappa/8$ \cite{Beffara2008}.
The inset of figure \ref{fig:w} displays the average length $\mathbb{E}\left[l\right]$ of the spanning perimeters versus the lattice system height $L_2 \in [32,1024]$; fitting gives $d_f = 1.355\pm0.01$ corresponding to the value $\kappa_{\mathrm{fractal}}$ given in table \ref{tab:kappas}.\\


\textit{Winding angle} -- How tortuous is an SLE can be exactly predicted: the winding angle measured between two typical points along the curve is Gaussian distributed, with a variance that grows logarithmically with the distance between the points \cite{Duplantier1988,Schramm2000,Wieland2003,Boffetta2008}.
On the lattice we define $\theta_j = \sum_{i=1}^j \alpha_i$, the sum of the local turns $\alpha_i$ that the curve takes at each step of its growth. In the scaling limit we expect that at distance $x$ along the curve from the starting point, the variance of $\theta$ is \cite{Boffetta2008,Saberi2009}:
\begin{equation}\label{eq:w}
    \mathrm{var}\left[\theta(x)\right] = a+\frac{2\kappa}{8+\kappa}\log x.
\end{equation}
The inset of figure \ref{fig:w} shows the rescaled distribution of $\theta$ at different distances $x$, falling on the Gaussian distribution, while the main plot shows its variance as function of $x$. Fitting according to (\ref{eq:w}) we obtain the estimation $\kappa_{\mathrm{winding}}$ of table \ref{tab:kappas}.\\


\textit{Left-Passage Probability} -- A famous result on SLE is the probability of the curve passing to the left of a given point in the upper-half plane. With the curve starting at the origin, this probability for a point $z=\rho e^{i\phi}$ depends only on $\phi$ and reads \cite{Schramm2001}
\begin{equation}\label{eq:LPP}
    P_\kappa(\phi) = \frac12 + \frac{\Gamma(\frac{4}{\kappa})}{\sqrt{\pi}\Gamma(\frac{8-\kappa}{2\kappa})}\cot(\phi)\,_2F_1\left[\frac12,\frac{4}{\kappa},\frac32,-\cot^2(\phi)\right]
\end{equation}
where $_2F_1$ is the ordinary hypergeometric function.
This prediction is seen to hold as well for dipolar curves not too far from their starting point, ie $\rho\ll L_y$ \cite{Bernard2007,Saberi2010}. We measure the left-passage probability $P(z)$ for a fixed set $\mathcal{S}$ of about 300 points in the semi-annulus $(\rho,\phi)\in[L_y/16, L_y/4]\times (0,\pi)$. We estimate $\kappa$ as the value minimising the mean-square deviation $Q(\kappa)$ \cite{Herrmann2012},
\begin{equation}\label{eq:Q}
    Q(\kappa)\equiv \frac{N-1}{\left|S\right|}\sum_{z\in S}\frac{\left[P(z)-P_\kappa(\phi(z))\right]^2}{P(z)\left(1-P(z)\right)}.
\end{equation}
$Q$ is plotted in the inset of figure \ref{fig:lpp}; minimising the interpolating function we find the estimate $\kappa_{\mathrm{LPP}}$ reported in table \ref{tab:kappas}. The main plot of figure \ref{fig:lpp} shows the data points for $P(z)$ averaged over $\rho$ for each value of $\phi$, together with the prediction $P_{\kappa_{\mathrm{LPP}}}$ of equation (\ref{eq:LPP}).

\textit{Critical exponents}-- Interfaces in RP being SLEs has consequences on the structural properties. It implies the existence of a so-called degenerate field $\Phi_{2,1}$ of dimension $h_{2,1} = (6-\kappa)/(2\kappa)$ \cite{Bauer2002,Cardy2005}, whose correlation functions satisfy differential equations \cite{BPZ84,Belavin1984}, leading in particular to relations involving the clusters' critical exponents \cite{Cardy2004,Doyon_2007}. We find that the previously measured RP fractal dimension, $D_f = 1.86(2)$ \cite{JacobsThorpe1995,mehdi2019} is indeed consistent with the prediction \cite{Janke2004} $D_f(\kappa) = 1+3/(2\kappa)+\kappa/8\sim 1.88$, where we used our value $\kappa_{RP}\sim2.9$. Moreover, there is also good agreement between the value of the correlation length exponent   $\nu=1.21(6)$ \cite{JacobsThorpe1995} and the expression \cite{Janke2004} $\nu(\kappa) = \left(2-2h_{2,1}\right)^{-1} = \kappa/(3\kappa-6)\sim1.1$. These two expressions give, for generic $\kappa\geq4$, the critical exponents of the $Q-$state Potts model geometric clusters \cite{Janke2004,Smirnov2007,Gamsa_2007}.\\

\textit{Conclusion}-- Using SLE theory, we first of all gave numerical evidence that the perimeters of spanning central-force RP clusters are SLE$_\kappa$ processes. We obtained independent estimates of the universal diffusion constant $\kappa_{\mathrm{RP}}\sim 2.9$, which by (\ref{eq:c}) corresponds to a central charge $c_{\mathrm{RP}}\sim 0.37$. These findings, along with the ones in \cite{nina2022} show that the rigorous approaches of SLE and CFT can be applied to the study of random geometry in a \emph{mechanical} percolation transition such as RP. Notably, SLE and CFT allow, by exploiting symmetry constraints, to express critical exponents solely in terms of $\kappa$, finding good numerical agreement with the values measured by standard techniques \cite{JacobsThorpe1995, mehdi2019}. Whether the value of $\kappa_{RP}$ corresponds to some simple fraction, which would lead to exact predictions for the RP critical exponents, remains a tantalizing possibility. 

At this stage it seems important to understand better the connection between RP and CP. Simulations of site- or bond-diluted RP show that a typical RP cluster consists  of overconstrained regions and isostatic (minimally rigid) dangling ends \cite{JacobsThorpe1995,Moukarzel1997,Moukarzel1999}. According to \cite{Liu2019},  clusters without overconstrained regions would fall in the CP universality class, so we expect that these latter drive the system to the distinct RP critical point. It would be therefore useful to tune the fraction of overconstrained regions (eg. cutting redundant bonds \cite{ellenbroek2015}), to interpolate between RP and MRP/CP, and analyze these transitions using the methods of the present work. One could also check if rigid hulls in MRP are indeed equivalent to CP hulls: in that case we expect correspondence with the percolation accessible perimeters \cite{AharonyBook} that are SLE$_{8/3}$ \cite{Duplantier2000,Beffara2004}.

More generally, our results open the way to applying the SLE analysis to other important classes of rigidity transitions besides central-force, providing a new tool to analyze in details the microstructure of disordered media at the onset of rigidity. A case particularly worth investigating is the frictional jamming transition \cite{silke2016,Liu2019}, where rigid clusters can be identified. It would be very interesting to examine if conformal invariance remains unbroken, and whether the critical exponents can be correctly predicted from SLE.



\begin{acknowledgments}
It is a pleasure to thank O.Abuzaid, F.Ares, M.Bouzid,  X.Cao, G.Delfino, N. El-Kazwini, D.X.Horvath, Y.Ikhlef, J.Jacobsen, S.Ribault, R.Santachiara and B.Walter for very valuable and stimulating discussions, as well as useful comments on the manuscript.
I acknowledge support from ERC under Consolidator grant number 771536 (NEMO).
\end{acknowledgments}


\end{document}